# Coupling of a surface plasmon with localized subwavelength microcavity modes

**P. Jouy[1], Y. Todorov[1,*], A. Vasanelli[1], R. Colombelli[2], I. Sagnes[3], and C. Sirtori[1]**

[1]Laboratoire ''Matériaux et Phénomènes Quantiques'', Université Paris Diderot-Paris 7, CNRS-UMR 7162, 75013 Paris, France

[2]Institut d'Electronique Fondamentale, Université Paris-Sud and CNRS-UMR 8622, F-91405 Orsay, France

[3]CNRS/LPN, Laboratoire de Photonique et de Nanostructures, Route de Nozay, 91460 Marcoussis, France

**Abstract :** Mid-infrared photonic modes of a periodically patterned metal-dielectric-metal structure have been investigated theoretically and experimentally. We have observed an anticrossing behaviour between cavity modes localized in the double-metal regions and the surface plasmon polariton, signature of a hybridisation between the two modes.

[*] Electronic address : yanko.todorov@univ-paris-diderot.fr

A metal-dielectric interface supports the well known surface plasmon polariton (SPP) modes, yielded by the coupling of the electromagnetic field with the coherent oscillations of the free electrons in the metal [1, 2]. Their ability to store and propagate the electromagnetic energy at sub-wavelength scales is essential for many applications such as bio-sensing, photonic circuits and optical data storage [3-5]. However, the SPP decay length in the air, $\delta$, strongly increases with the wavelength $\lambda$, compromising the subwavelength energy confinement in the mid-infrared (mid-IR) and THz part of the spectrum [6]. Recently, the concept of "spoof" surface plasmons has been put forward [7], where shorter decay lengths can be engineered by drilling subwavelength holes in a thick metal substrate. An older, but efficient method of localizing the SPP near the surface is coating the metal with a thin high refractive index dielectric material [8]. In this letter we report on structures where a metallic strip grating is deposited on a GaAs-coated metal. This structure supports modes which are localized between the two metals [9,10], and yet able to efficiently couple with the surface plasmon guided by the air-semiconductor-metal multilayer. This coupling effect occurs for well defined choices of the grating periodicity and GaAs thickness and allows to control the properties of the free propagating surface waves.

In our experimental structure a 300 nm thick undoped semiconductor (Gallium Arsenide) slab is sandwiched between two golden parts, one of which is patterned as a rectangular strip grating of width $s$ and period $p$, as illustrated in Fig. 1(a). The typical dimensions of the structure are $s = 0.4$-$3.5\mu$m and $p = 1.5$-$4.8\mu$m and have been fabricated by e-beam lithography. The thickness of the grating strips is 87nm. Similar structures, but with micrometric dimensions, are known to support electromagnetic modes which are strongly localized under the grating strips, in the microwave and THz regions [9,10]. These modes originate from the $TM_0$ mode confined in the double metal regions [10]. The multiple reflections of the $TM_0$ mode between the two ends of the metal strips yield standing wave patterns, as illustrated in Fig.1(b). The resonant wavelengths $\lambda_K$ are provided by the formula:

$$\lambda_K = 2n_M(\lambda_K)s / K \tag{1}$$

Here $K$ is a non-zero integer, corresponding to the number of nodes of the vertical electric field $E_z$ and $n_M$ is an effective index which depends on the metal loss and the impedance mismatch on the resonator ends [10]. This quantity is wavelength dependent $n_M=n_M(\lambda)$, varying monotonically between 4.5 and 6.6 in the explored wavelength range ($\lambda= 5$-$15\mu$m).

As expected, these values are higher than the refractive index of the semiconductor, $n_{GaAs} = 3.3$ [10].

The structures are analysed by performing reflectivity spectra, where the cavity modes appear as lorentzian dips of typical full width at half maximum of 25meV. The corresponding typical quality factors are $Q\sim6$, lower than the values $Q\sim10$ observed in the far-infrared region [10], due to the increased metal losses at shorter wavelengths. Fig. 1(c) summarizes the experimentally measured resonant wavelengths of the first (*K=1*) and the second (*K=2*) order, for an incident angle of $\theta = 45°$. The measured effective index is $n_M \sim 4.6$ for the first order (*K=1*) resonance. For these structures the grating period is kept low, $p < 3\mu m$, and it remains essentially smaller than the wavelength, as in Ref. 10. For these periods the cavity resonances are below the onset of diffraction, and they do not feature any dispersion as a function of the incident wavector $k_x = (\omega/c)\sin\theta$.

However, the situation changes completely when the period $p$ is comparable to $\lambda/2$. In this case the diffraction phenomena interfere with the localised modes and modify their line shape and dispersion relation. The first diffraction order to come into play is the -1 diffracted order, which appears for a wavelength $\lambda_{-1}$:

$$\lambda_{-1} = p(1 + \sin\theta) \qquad (2)$$

This is illustrated in the data set presented in Figures 2(a) and (b), in (red) continuous curves. For all samples, the strip width is $s = 0.8\mu m$, which sets the energy of the fundamental cavity mode (*K=1*) at 151 meV ($\lambda_{K=1}$=8.2μm). We have then let vary either the angle of incidence $\theta$, or the grating period $p$, in order to tune the diffraction edge (Eq.(2)) across the cavity resonance. In Fig. 2(a), the incident angle is fixed, $\theta$=45°, and the reflectivity is studied for progressively increasing grating periods $p$. We observe that when $\lambda_{K=1} = \lambda_{-1}$ the cavity mode splits in two. Similar behavior is observed from the data reported in Fig.2(b) where the period is fixed, $p$=4.8μm, but the incident angle $\theta$ is varied. For low angles we recover the "bare" cavity mode at 151meV, and the cavity mode splits in two for large incident angles, when $\lambda_{K=1}=\lambda_{-1}$.

The experimental reflectivity spectra of Fig. 2(a) (b) are very well reproduced by our simulations, indicated by (black) dotted curves, based on the modal method formalism with surface impedance boundary conditions [11]. The input parameter for the model is the gold dielectric constant, for which we use a Drude-like susceptibility $\varepsilon(E) = 1 - 2.29 \times 10^7 / \left[ E(E + i145) \right]$, with $E$ the photon energy in meV. Note that the metal loss in this formula is increased with respect to the bulk values known from the literature [12], possibly due to scattering on the semiconductor-metal interface and the presence of thin (6 nm) Ti adhesive layers between the semiconductor and the gold coatings.

The effect of the diffraction edge on the cavity resonance is best illustrated in Fig. 3(a) where we plot the simulated reflectivity as a function of the photon energy $E$ and incident wavector $k_x$, for the 1$^{st}$ Brillouin zone of the grating with $p$ = 4.8µm, together with the experimental peak positions from Fig. 2(b) ((red) dots). This diagram shows that the cavity mode at 151meV is indeed dispersionless far from the diffraction limit, whereas a clear anti-crossing behaviour is observed close to the -1 diffraction line. This anti-crossing appears because of the coupling between the cavity mode and the surface plasmon-polariton (SPP) supported by the air-semiconductor-metal multilayer, excited in the vicinity of $\lambda_{-1}$ [2]. Note that, away from the anti-crossing region, our structure acts like a leaky wave antenna for the SPP guided in the multilayer [13]. The dispersion relation of this SPP is plotted in Figure 3(b), for the case of planar GaAs-coated metal surface without any grating. For the sake of comparison with Fig.3(a) the wavector is still normalized to $\pi/p$ and the energy of the bare cavity mode is indicated with a dashed line. From this diagram, we deduce an anti-crossing wavector $k_x$=1.22$\pi/p$, which corresponds to a (2-1.22)$\pi/p$=0.78$\pi/p$ in the 1$^{st}$ Brillouin zone of the grating, in accordance with the experimental results reported in Fig.3(a).

The dispersion relation of the GaAs-coated plasmon in Fig.3(b) gives a hint on the coupling mechanism. Indeed, for mid-infrared frequencies, the simple air-metal surface plasmon follows tightly the light line, whereas the 300nm GaAs-coated plasmon is increasingly pulled below the air light line for higher energies (shorter wavelengths). The SPP decay length in the air can be simply expressed as $\delta = 1 / \sqrt{k_x^2 - \omega^2 / c^2}$ and therefore for higher photon energies, while $\delta$ decreases the SPP has a stronger overlap with the semiconductor region, favoring the interaction with the strongly localized cavity modes. This is indeed the case as illustrated in Figure 4(a), which resumes the simulated mode splitting as a function of the coating thickness $L$. For these simulations, the grating period is $p$=4.8µm, and for convenience the metal strip width is chosen to be $s$=0.6µm (cavity resonance at $\lambda_{K=1}$=6.6µm), so that the anti-crossing appears in the middle of the 1$^{st}$ Brillouin zone of the grating, at $k$=0.5π/$p$. In Figure 4(b) we have plotted $\delta$ (open squares) of the planar GaAs-coated surface plasmon as well as the electric field overlap with the semiconductor region (triangles), as a function of $L$, for the wavevector corresponding to the anti-crossing point. As expected, for increasing thickness the overlap of the plasmon field with the semiconductor also increases, and $\delta$ decreases. Note that even subwavelength coatings ($L$=400nm) are sufficient to reduce significantly the plasmon decay length in the air: from 120µm to 20µm.

The strong interaction between the surface plasmon and the cavity modes means a periodic exchange of the electromagnetic energy between the two. As shown in Ref. 10 the cavity modes have the ability to convert and store very efficiently the energy of the incoming wave into evanescent near field. Therefore, thanks to the coupling mechanism evidenced above, the cavity modes mediate the energy transfer between the radiation modes and the evanescent surface plasmon mode, allowing the excitation of the SPP with an increased efficiency. The interference between the two coupled modes with the evanescent background of the grating also explains the distorted Fano-like line-shapes observed in the spectra of Figures 2(a) and 2(b) close to the diffraction line, as pointed out by Sarrazin and al. [14].

In conclusion, we have experimentally demonstrated strongly localized cavity modes supported by compact metal-semiconductor-metal grating structures scaled to operate at the mid-IR frequency range. We have observed the strong coupling between these modes and the surface plasmon also supported by the air-dielectric-metal structure. The latter has an increased confinement of the electromagnetic energy above the metal-air interface. Since for this frequency range the semiconductor can be engineered into a quantum cascade gain medium, this opens a variety of perspectives for electrical generation or amplification [15] of surface plasmons, and the building of very compact SPP circuits in the mid-IR frequency range.

We thank Michael Rosticher (Ecole Nomale Supérieure, Paris) for help with the e-beam lithography system. This work has been partially supported by the French National Research Agency (ANR) in the frame of its Nanotechnology and Nanosystems program P2N, project ANR-09-NANO-007 and the ERC grant "ADEQUATE". Part of the device fabrication has been performed at the CTU-IEF-Minerve which was partially funded by the "Conseil Général de l'Essonne".

**Figure captions**

Fig.1. (a) Scanning electron microscope picture of the device in false colours, with the relevant geometrical parameters. (b) Simulation of the vertical electric field ($E_z$) distribution for the lowest order cavity mode. (c) Resonant wavelengths for the first order ((red) dots) and second order ((blue) squares) as a function of the strip widths $s$, determined from reflectivity measurements.

Fig. 2. (a) Experimental reflectivity spectra (continuous (red) lines) for structures with strip width $s$=0.8µm and variable grating period $p$. The incident angle is $\theta$=45°. (b) Reflectivity spectra (continuous (red) lines) for a structure with strip width $s$ = 0.8µm and period $p$=4.8µm, for different incident angles $\theta$. In both (a) and (b) we provide simulated spectra in (black) dotted curves. The dashed line indicates the bare cavity resonance.

Fig. 3. (a) Simulated reflectivity as a function of the photon energy $E$ and wavevector $k_x$, for the 1$^{st}$ Brillouin zone of the grating with $p$ = 4.8µm, together with the experimental peak positions from Fig. 2(b) ((red) dots). (b) Dispersion relation of the GaAs coated metal planar surface plasmon. The wavector $k_x$ has been normalised to the grating wavevector $\pi/p$ ($p$ = 4.8µm) for comparison with (a). The dashed line indicates the bare cavity resonance.

Fig 4. (a) Mode splitting at the anti-crossing point as a function of the GaAs thickness $L$, computed for a structure with $p$=4.8µm and $s$=0.6µm. For these simulations, the anti-crossing appears for a wavevector $k_x$ =0.5$\pi/p$. (b) Decay length (open squares) and vertical electric field ($E_z$) overlap with the semiconductor layer (triangles) for a planar surface plasmon, as a function of the GaAs thickness $L$.

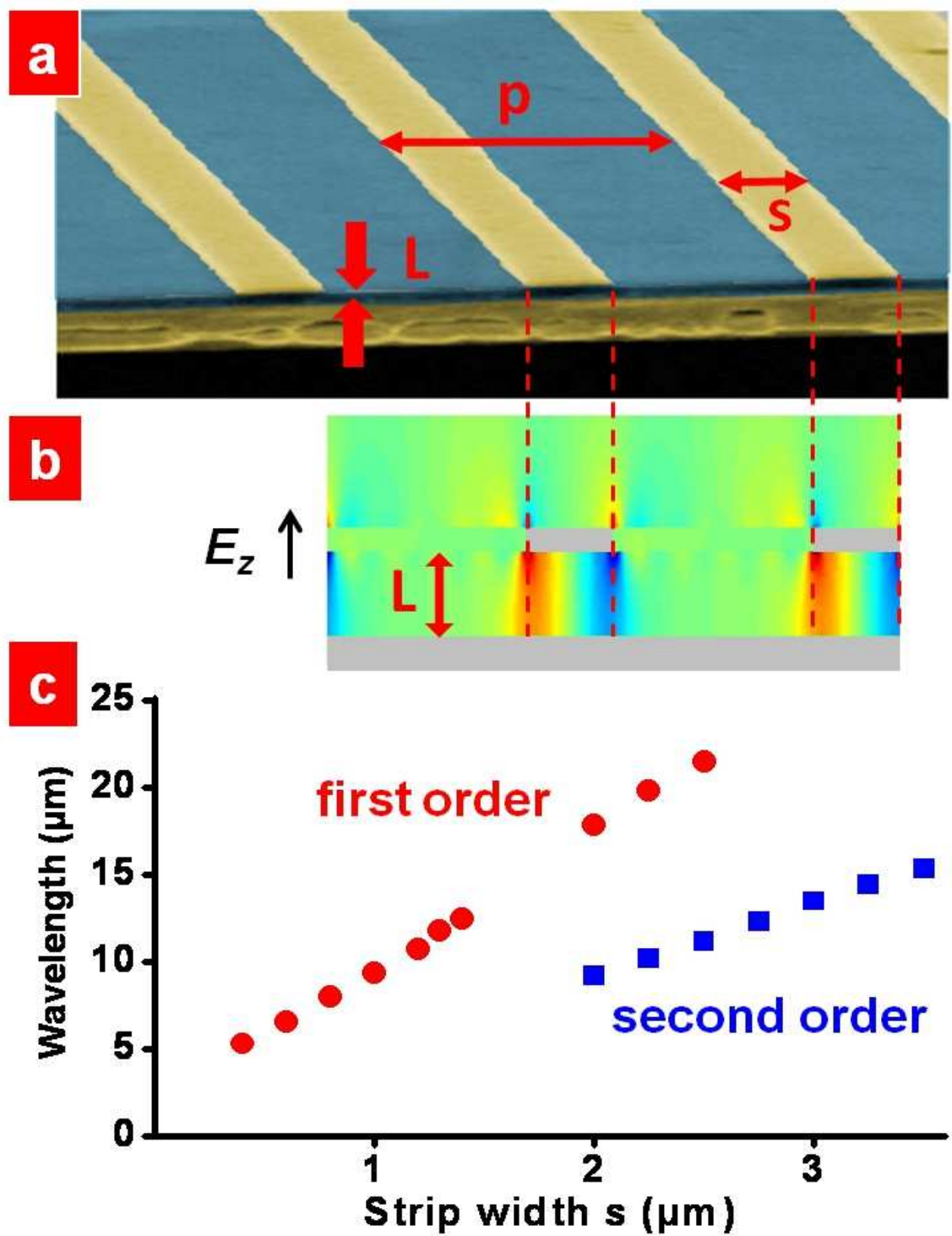

Figure 1.

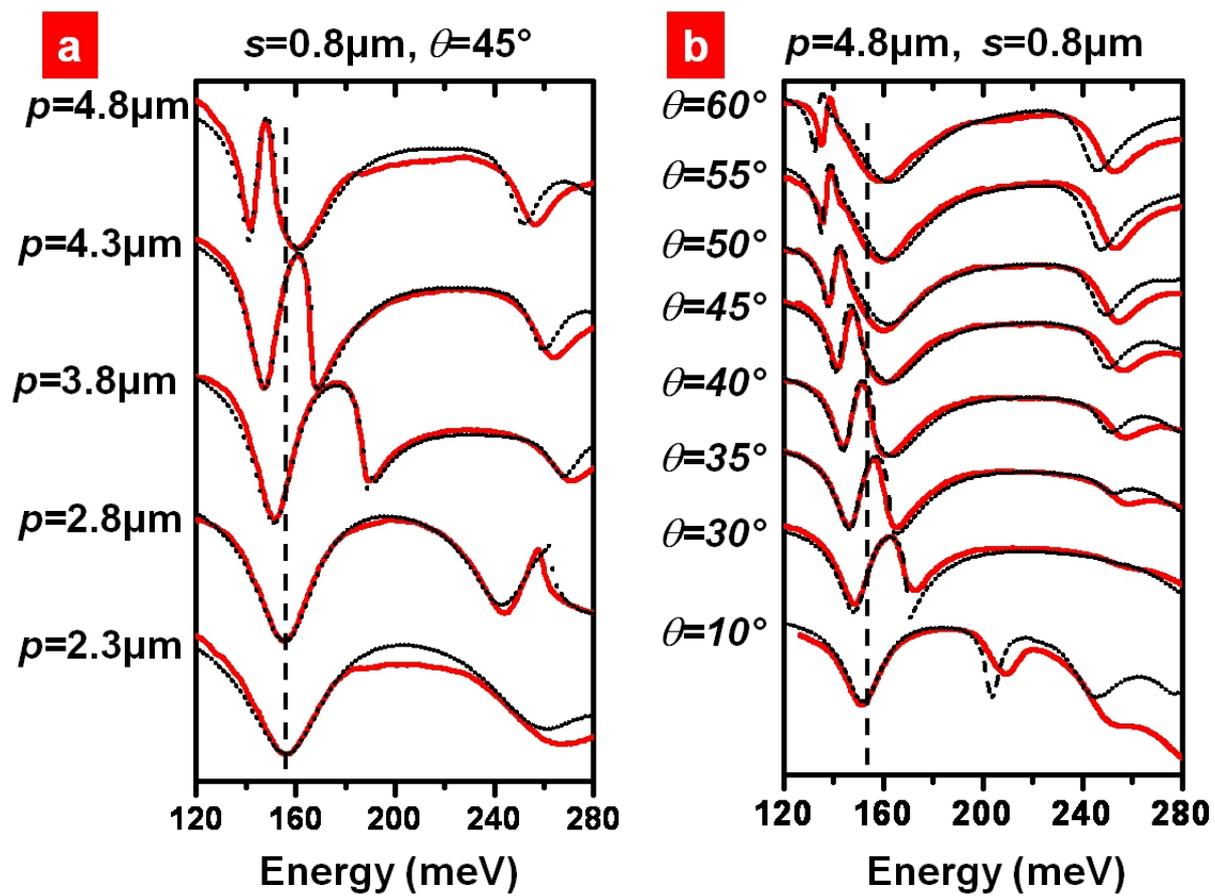

Figure 2.

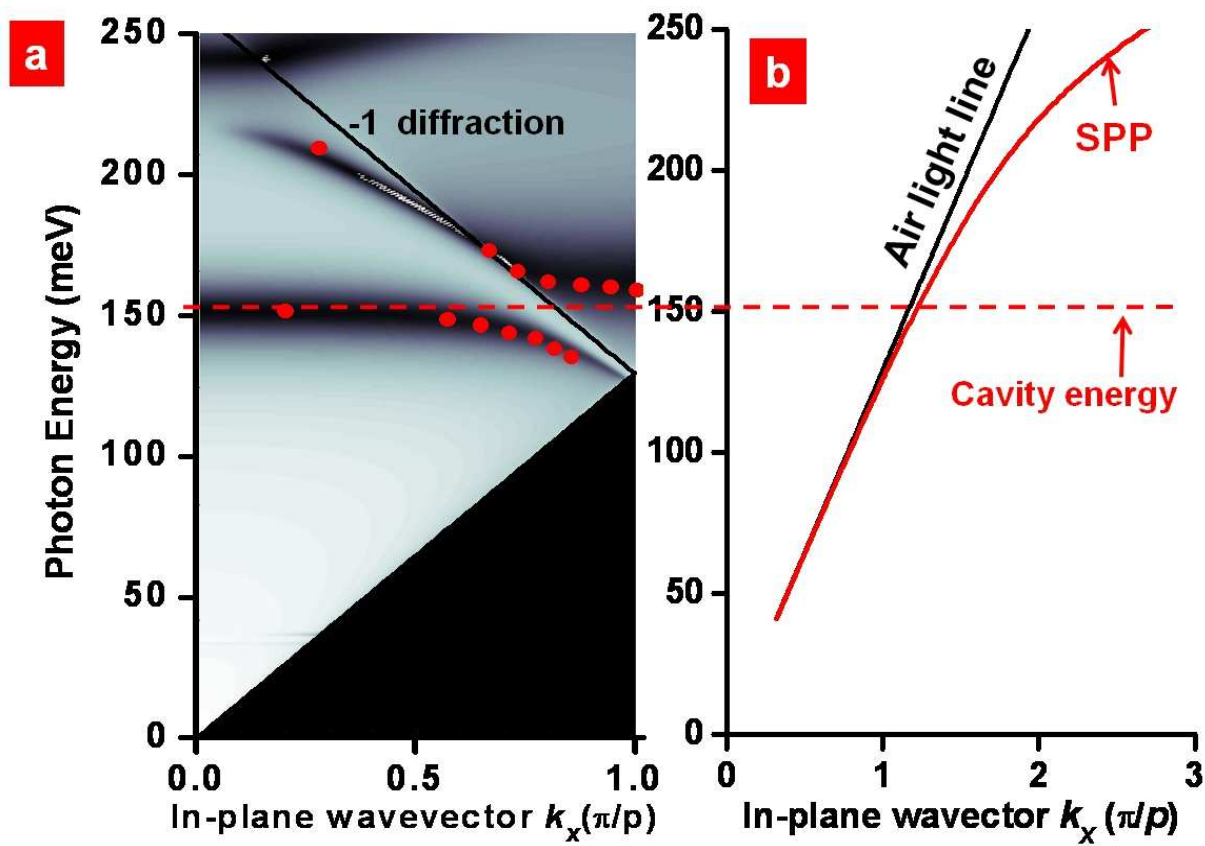

Figure 3.

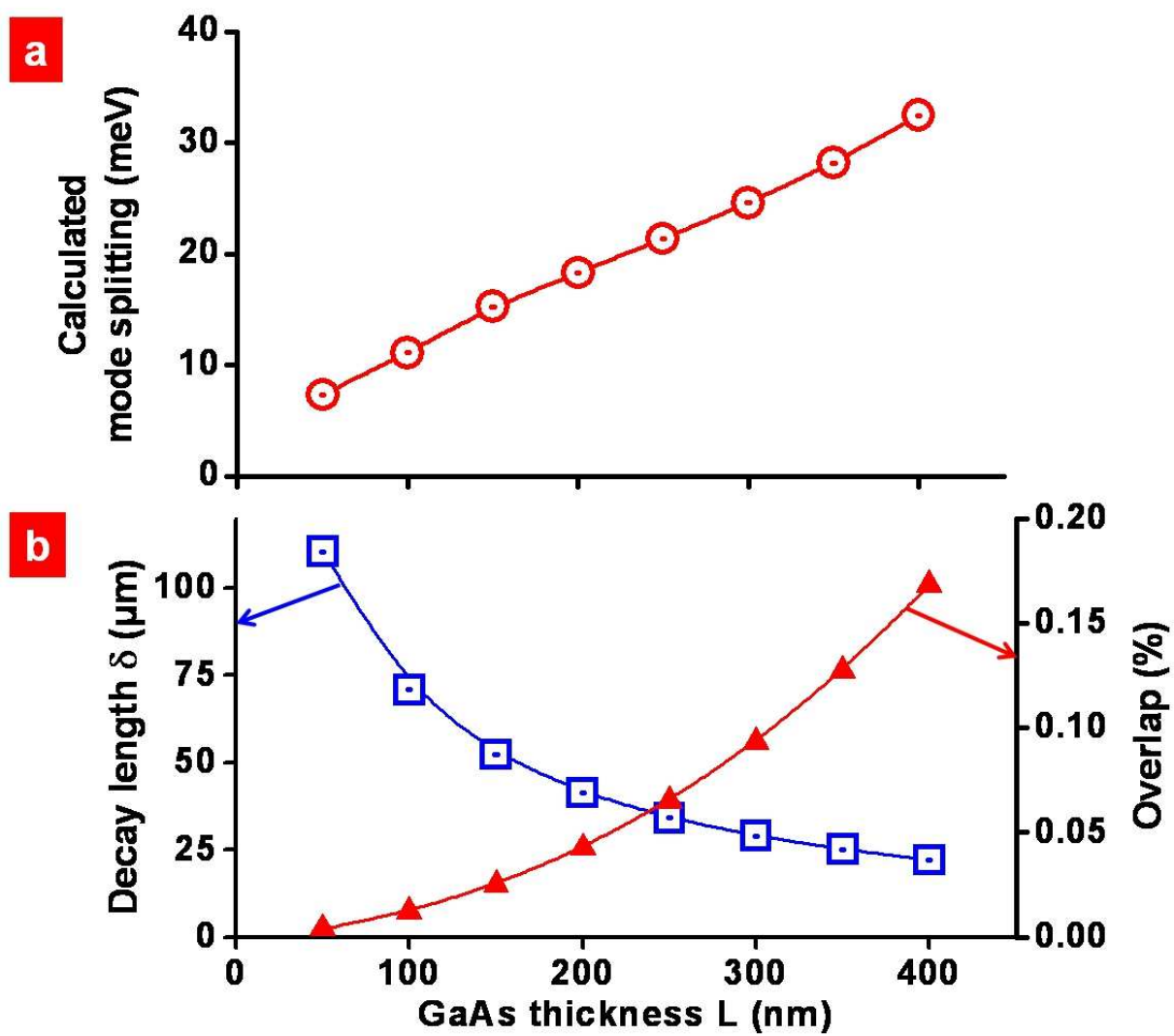

Figure 4.